\documentclass{PoS}

\newcommand{\ba}{\begin{eqnarray}}
\newcommand{\ea}{\end{eqnarray}}
\newcommand{\bd}{\begin{displaymath}}
\newcommand{\ed}{\end{displaymath}}
\newcommand{\be}{\begin{equation}}
\newcommand{\ee}{\end{equation}}
\newcommand{\mev}{\mathrm{MeV}}

\newcommand{\fm}{\mathrm{fm}}
\newcommand{\atwo}{a^{I=2}_{\pi\pi}}
\newcommand{\plotangle}{270}
\newcommand{\mywidth}{162pt}
\newcommand{\chipt}{$\chi$PT}
\newcommand{\fphy}{f_{\pi,\mathrm{phy}}}
\newcommand{\Ref}[1]{Ref.~#1}

\newcommand{\E}{E^{I=2}_{\pi\pi}}
\newcommand{\dE}{\delta E^{I=2}_{\pi\pi}}
\newcommand{\Fig}[1]{Fig.~#1}
\newcommand{\Eq}[1]{Eq.~#1}
\newcommand{\gap}{\hspace{10pt}}

\title{Scattering from finite size methods in lattice QCD}

\ShortTitle{Scattering from finite size methods in lattice QCD}

\author{  \begin{flushright}
          DESY 09-171\\
          SFB/CPP-09-95\\
          MS-TP-09-21
          \end{flushright}
        \speaker{Xu Feng}\,\,$^{a,b}$,
        Karl Jansen$^a$ and Dru B. Renner$^a$\\
        \llap{$^a$}NIC, DESY, Platanenallee 6, D-15738 Zeuthen, Germany\\
        \llap{$^b$}Universit\"at M\"unster, Institut f\"ur Theoretische Physik, Wilhelm-Klemm-Strasse 9, D-48149 M\"unster, Germany\\
        E-mail: \email{xu.feng@desy.de}}

\abstract{Using two flavors of maximally twisted mass fermions, we
calculate the S-wave pion-pion scattering length in the isospin
$I=2$ channel and the P-wave pion-pion scattering phase in the
isospin $I=1$ channel. In the former channel, the lattice
calculations are performed at pion masses ranging from $270~\mev$ to
$485~\mev$. We use chiral perturbation theory at next-to-leading
order to extrapolate our results. At the physical pion mass, we find
$m_\pi \atwo=-0.04385\,(28)(38)$ for the scattering length. In the
latter channel, the calculation is currently performed at a single
pion mass of $391~\mev$. Making use of finite size methods, we
evaluate the scattering phase in both the center of mass frame and
the moving frame. The effective range formula is employed to fit our
results, from which the rho resonance mass and decay width are
evaluated.
 \vspace{40pt}
\begin{center}
\includegraphics[width=100pt]{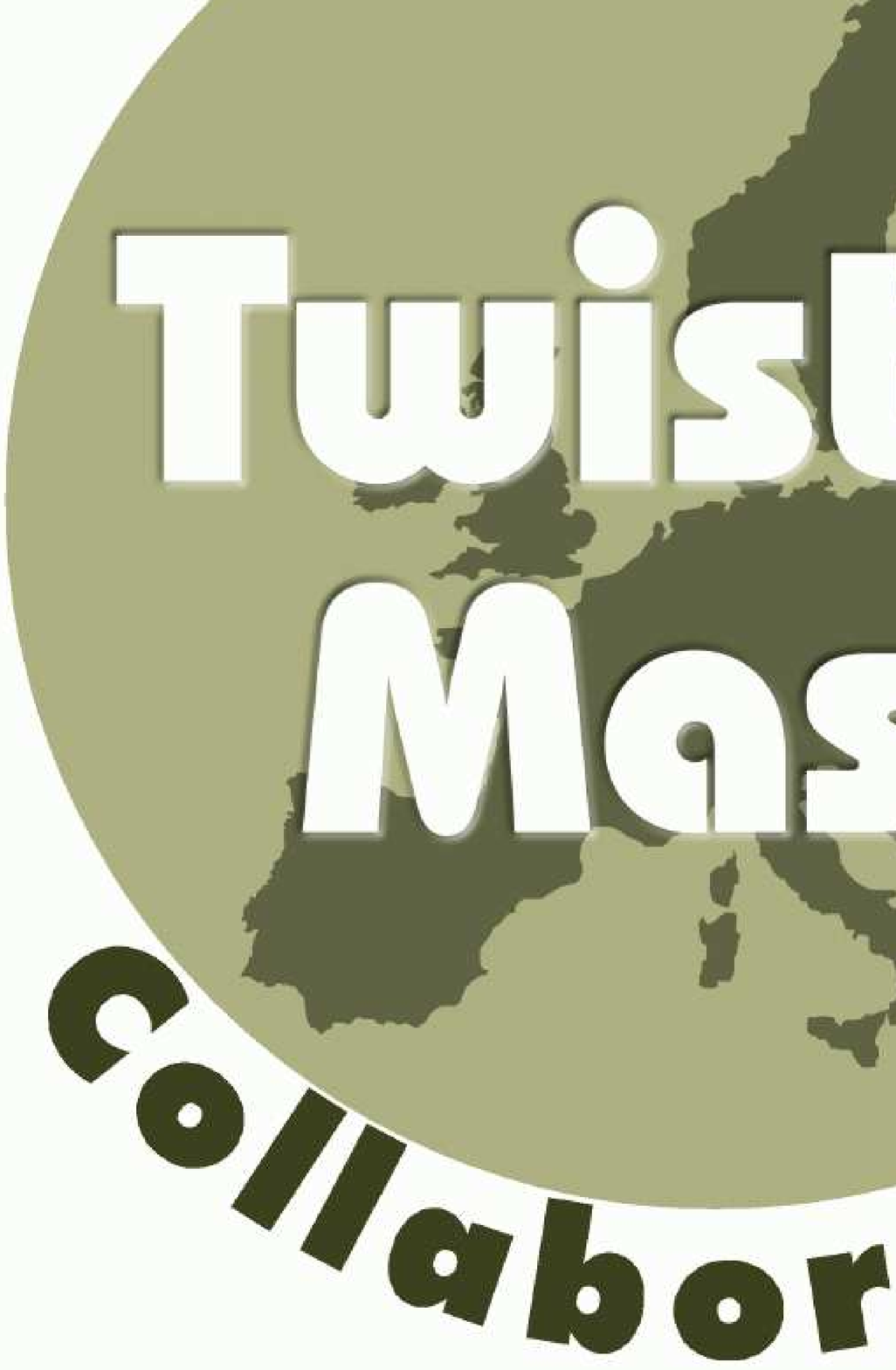}
\end{center}
}

\FullConference{The XXVII International Symposium on Lattice Field Theory - LAT2009\\
         July 26-31 2009\\
         Peking University, Beijing, China}

\begin{document}
\section{Introduction}

Experimentally, hadron-hadron scattering is an important method to
study the strong interactions. Among the various scattering
possibilities, pion-pion scattering is the simplest and best
understood one due to the fact that the underlying chiral symmetry
strongly determines the low energy behavior of the pion-pion
scattering amplitude. Despite its simplicity, pion-pion scattering
offers us a lot of information on the strong interactions. In the
isospin $I=2$ channel, near threshold the S-wave scattering length
is used to determine the corresponding low energy constants (LECs)
of chiral perturbation theory (\chipt). In the $I=1$ channel, the
prominence of the rho resonance is clearly observed. By measuring
the P-wave scattering phase, the parameters for the resonance mass
and decay width can be extracted. In the $I=0$ channel, the sigma
resonance appears in pion-pion scattering. In contrast to the rho
resonance, a precise identification of the sigma resonance remains a
great challenge because the large decay width of the sigma causes a
strong overlap between it and its background.

Pion-pion scattering is non-perturbative in nature at low energies.
Therefore, it should be studied with a non-perturbative method like
lattice QCD. In the center-of-mass frame (CMF), a direct lattice QCD
determination of the scattering phase is possible by employing
L\"uscher's finite-size
methods~\cite{Luscher:1985dn,Luscher:1986pf,Luscher:1990ck,Luscher:1990ux,Luscher:1991cf},
which establish relations between the discrete energy spectrum in
the finite volume and the elastic scattering phase in the infinite
volume. In the moving frame (MF), where the total momentum of the
pion-pion scattering system is fixed to be a non-zero value, one can
evaluate the scattering phase by using the method proposed by
Rummukainen and Gottlieb~\cite{Gottlieb:1995dk}, which is an
extension of L\"uscher's method to MFs. To perform our calculations,
we use the $N_f=2$ maximally twisted mass fermion ensembles from the
European Twisted Mass Collaboration (ETMC). Due to the properties of
twisted mass fermions at maximal twist, our calculation is
automatically accurate to $O(a^2)$ in the lattice spacing, $a$.

In this paper, we present a calculation of the S-wave pion-pion
scattering length in the $I=2$ channel and the P-wave scattering
phase in the $I=1$ channel. A calculation of pion-pion scattering in
the $I=0$ channel using $2+1$ flavors of domain wall fermions has been
reported recently~\cite{Liu:2009uw}. Although the object of our
investigation is simply pion-pion scattering, the approach to study
scattering from finite size methods in lattice QCD is universal and
can be applied to other meson-meson, meson-baryon and baryon-baryon
scattering systems.

\section{$I=2$ channel}
\label{sect:I_2} In the $I=2$ channel, the lattice calculation is
performed in the CMF. As mentioned in the introduction, L\"uscher's
finite size method relates the energy levels of two pion states in a
finite volume to the scattering phase in the infinite volume.  For
the case of two pions with zero relative three-momentum, this method
establishes a relationship between the lowest energy eigenvalue
$E_{\pi\pi}^{I=2}$ in a finite box of size $L$ and the corresponding
scattering length $\atwo$~\cite{Luscher:1986pf}:
\bd \dE = \E - 2m_\pi = - \frac{4\pi\atwo}{m_\pi L^3} \left[ 1 +
c_1\frac{\atwo}{L} + c_2\left(\frac{\atwo}{L}\right)^2 \right] +
O(L^{-6})\,, \ed
where $c_1=-2.837297$ and $c_2=6.375183$ are numerical constants.
Thus the above result allows us to convert a lattice determination
of the energy shift $\dE$ into a calculation of $\atwo$.

\begin{figure}[htb]
\begin{minipage}{210pt}
\hspace{0pt}\includegraphics[width=\mywidth,angle=\plotangle]{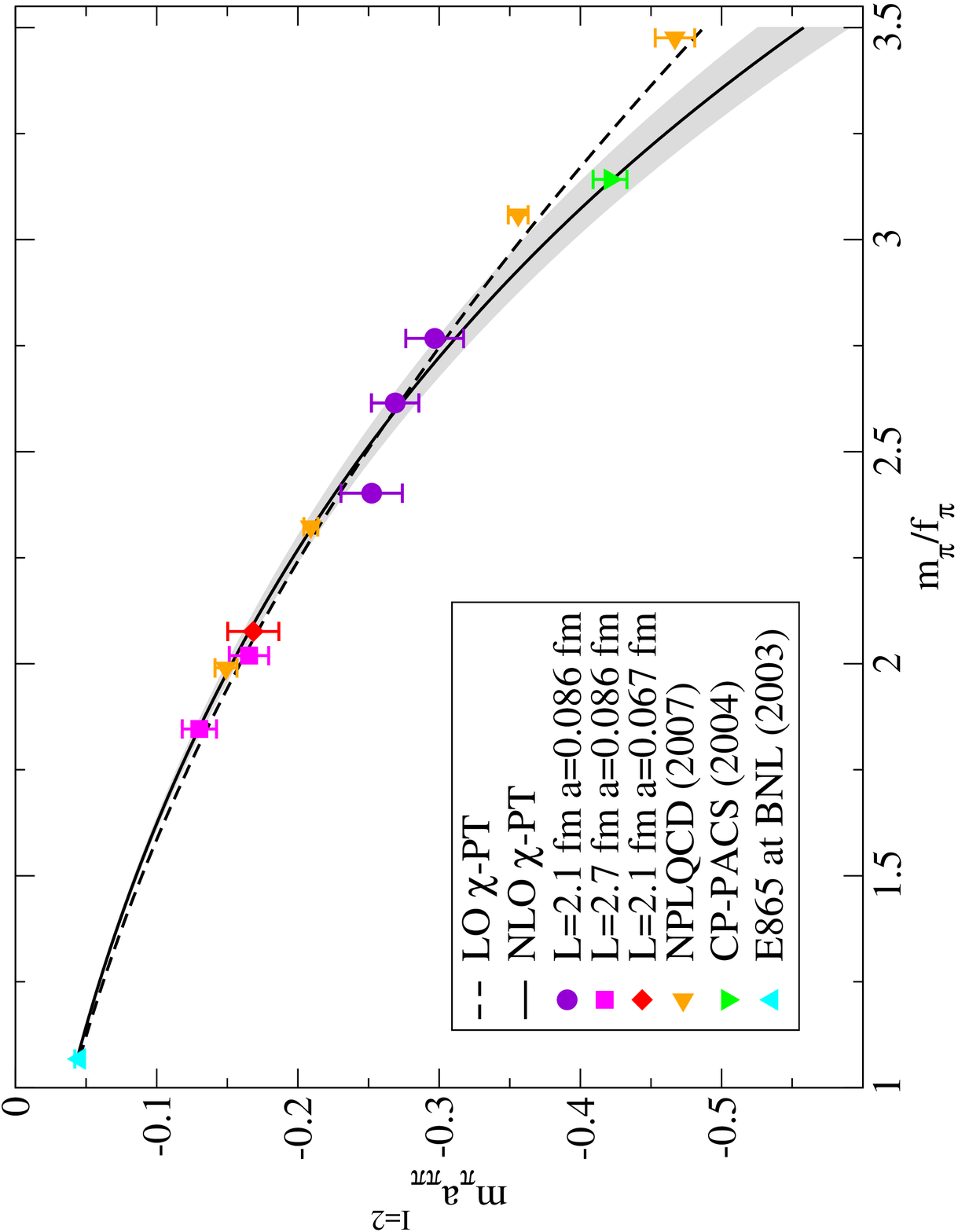}
\caption{Chiral extrapolation for the I=2 pion-pion scattering
length. The results in this work are shown together with the lattice
calculations of NPLQCD~\cite{Beane:2005rj,Beane:2007xs} and
CP-PACS~\cite{Yamazaki:2004qb} and the experimental data from E865
at BNL~\cite{Pislak:2003sv}.} \label{fig:fit}
\end{minipage}
\hspace{4pt}
\begin{minipage}{210pt}
\hspace{0pt}\includegraphics[width=\mywidth,angle=\plotangle]{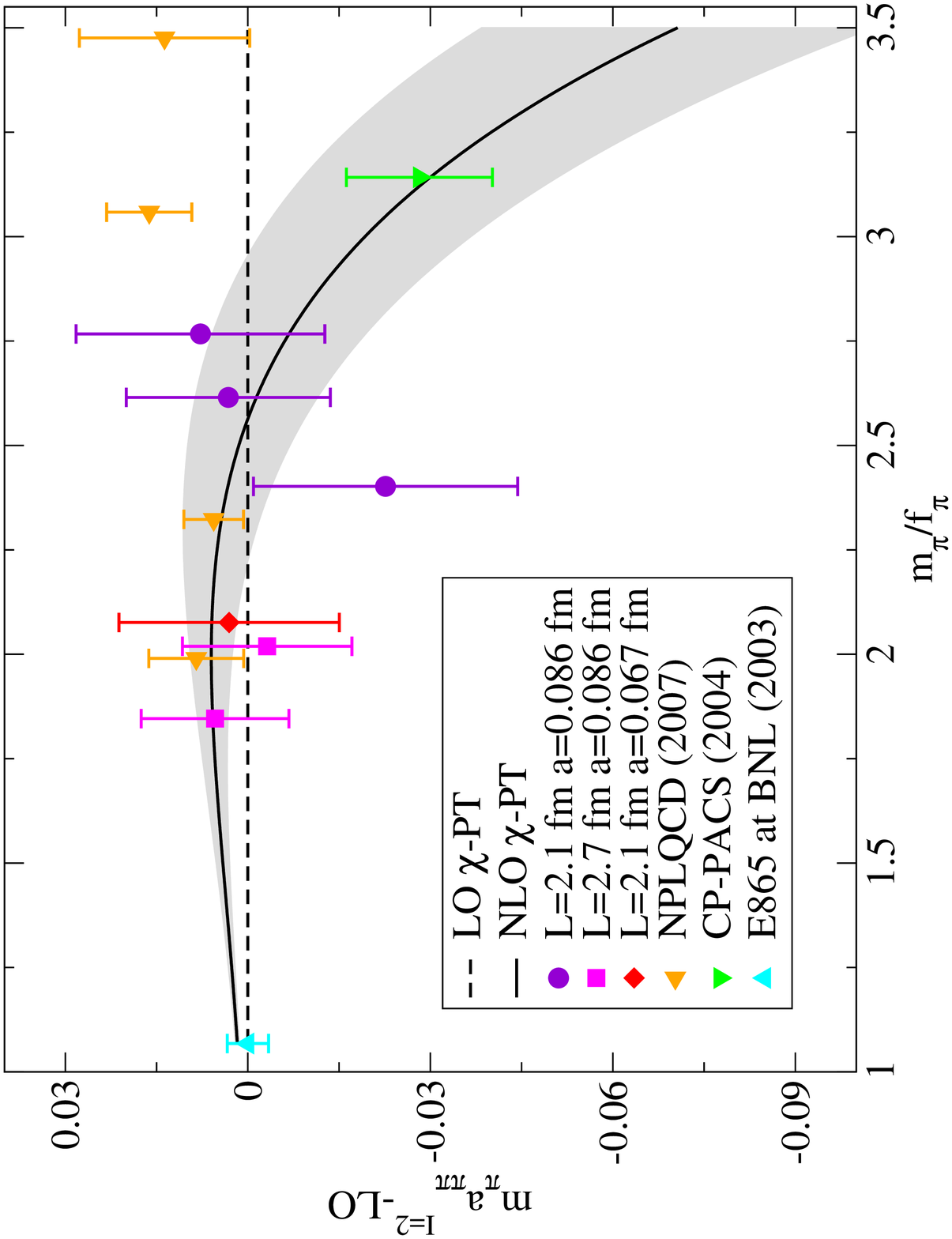}
\caption{Difference between the lattice calculation of the
scattering lengths and the LO {\chipt} prediction. The scattering
lengths agree statistically with the LO {\chipt} prediction for pion
masses ranging from $m_\pi =270~\mev$ to $485~\mev$.}
\label{fig:fit1}
\end{minipage}
\end{figure}

To extract the energy shift, $\dE$, and then the scattering length,
$\atwo$, we use the two flavor maximally twisted mass fermion
configurations from
ETMC~\cite{Boucaud:2007uk,Dimopoulos:2008sy,Urbach:2007rt}.\footnote{The
role of the neutral pion in our setup is discussed in some detail in
\Ref{\cite{Dimopoulos:2009qv}}} The pion masses range from
$m_\pi=270~\mev$ to $485~\mev$. For most of the ensembles, the
lattice spacing is $a=0.086~\fm$ and the box size is $L=2.1~\fm$.
For the lower pion masses the volume is increased to $L=2.7~\fm$.
Additionally, we perform a check for lattice artifacts with a single
calculation at a finer lattice spacing of $a=0.067~\fm$. All the
results for the scattering length are shown in \Fig{\ref{fig:fit}}.

The next step is to extrapolate the scattering length to the
physical limit. Here, we make use of NLO {\chipt} for the pion-pion
scattering length, which has recently been studied in the twisted
mass case~\cite{Buchoff:2008hh}.
%
%
The {\chipt} fit curves are shown in \Fig{\ref{fig:fit}}. In the
same figure, we also provide a comparison to the lattice results of
NPLQCD~\cite{Beane:2005rj,Beane:2007xs} and
CP-PACS~\cite{Yamazaki:2004qb} and the experimental data from E865
at BNL~\cite{Pislak:2003sv}.  We find general agreement between our
calculation and the results of NPLQCD at similar pion masses.
Additionally, we find agreement with the experimental determination
of $m_\pi\atwo$. To highlight the impact of the NLO terms in the
{\chipt} description of the pion mass dependence of $m_\pi\atwo$, we
show the difference between the lattice calculations of the
scattering length and the LO {\chipt} prediction in
\Fig{\ref{fig:fit1}}. We find that the scattering lengths
statistically agree with the LO {\chipt} result for all lattice
calculations with $m_\pi < 500~\mev$. Accordingly, the NLO {\chipt}
functional form provides a reasonable description of the lattice
results in the same region of $m_\pi$. At the physical pion mass, we
obtain the final results
\bd m_\pi \atwo=-0.04385\,(28)(38) \gap\textmd{and}\gap
l_{\pi\pi}^{I=2}(\mu=\fphy)=4.65\,(.85)(1.07)\,, \ed
where the first error is statistical and the second is our estimate
of several systematic effects. For more details, we refer the reader
to our recent paper~\cite{Feng:2009ij}.
\section{$I=1$ channel}
\label{sect:I_1} In the $I=1$ channel, the rho meson decays into two
pions in the P-wave. As the case of S-wave, one can make use of
finite size methods to calculate the P-wave scattering phase.
However, the extraction of the rho resonance from the scattering
phase is non-trivial for several reasons. First, only when the pion
masses are light enough to satisfy the requirement of
$m_\pi<m_\rho/2$, is it possible for the rho to decay into two
pions. Second, the standard form of L\"uscher's method is derived to
address the elastic scattering process, so the interesting energy
spectrum should be smaller than $4m_\pi$ to avoid the inelastic
scattering. Third, because of the finite volume, the energy spectrum
of pion-pion scattering states is discrete, which translates into
scattering phases at discrete energies. Therefore, an analytic
expression of the scattering phase is required to describe its
dependence on the energy spectrum. Usually, one employs the
effective range formula to meet this demand:
    \bd
    \label{eq:effective_range}
    \tan{\delta_1}(k)=\frac{g^2_{\rho\pi\pi}}{6\pi}\frac{k^3}{E_{CM}(M_R^2-E_{CM}^2)}
    \;,\quad k=\sqrt{E_{CM}^2/4-m_\pi^2}\;,
    \ed
where $\delta_1(k)$ is a P-wave scattering phase in the $I=1$
channel and $E_{CM}$ is the center-of-mass energy. In the MF,
$E_{CM}$ is simply given by $E_{CM}^2=E^2-\vec{P}^2$, where $E$ is
the discrete energy eigenvalue and $\vec{P}$ is the total momentum
of the MF.\footnote{To reduce lattice discretization effects, we use
the relation $\cosh(E_{CM})=\cosh(E)-2\sin^2(P/2)$ instead.} Thus,
in the effective range formula, only two parameters are
undetermined, $M_R$ and $g_{\rho\pi\pi}$, where $M_R$ denotes the
resonance mass and $g_{\rho\pi\pi}$ is the effective
$\rho\rightarrow\pi\pi$ coupling constant, which largely determines
the size of resonance decay width:
    \bd
    \Gamma_R=\frac{g^2_{\rho\pi\pi}}{6\pi}\frac{k^3}{M_R^2}
    \;,\quad k=\sqrt{M_R^2/4-m_\pi^2}\;.
    \ed
By fitting the discrete scattering phases to the effective range
formula, one can evaluate the parameters $M_R$ and $g_{\rho\pi\pi}$
and then determine $\Gamma_R$. Conversely, by using the latest
PDG~\cite{Amsler:2008zzb} values of $m_{\pi}=139.5702(4)~\mev$,
$M_\rho=775.49(34)~\mev$ and $\Gamma_\rho=149.1(8)~\mev$, one can
also evaluate $g_{\rho\pi\pi}=5.98(2)$ at the physical pion mass.

With the effort required to simulate with light up and down quark
masses, the condition of $m_\pi<m_\rho/2$ has been satisfied by only
a few lattice
calculations~\cite{Aoki:2007rd,Gockeler:2008kc,Jansen:2009hr}. So
far, all these studies concentrated on one or two scattering phases
for each ensemble. Since the effective range formula carries two
unknown parameters, more scattering phases are needed for a precise
fit. To accomplish this goal, a natural way is to calculate the
energy spectrum of the higher excited states. In order to isolate
the ground state and the first excited state, we set up a $2\times2$
correlation matrix
     \ba
     \label{eq:correltaion_matrix}
     C_{2\times2}(t)=\left(
     \begin{array}{cc}
     \langle (\pi\pi)^\dagger(t)(\pi\pi)(0)\rangle
     & \langle (\pi\pi)^\dagger(t)\rho(0)\rangle\\
     \langle\rho^\dagger(t)(\pi\pi)(0)\rangle
     & \langle\rho^\dagger(t)\rho(0)\rangle \\
     \end{array}
     \right)\;,
     \ea
where the interpolating operator $(\pi\pi)(t)$ has the same quantum
numbers, $J^{PC}=1^{--}$, as the interpolating operator $\rho(t)$.
After diagonalization of the matrix in~\ref{eq:correltaion_matrix},
we obtain the energy eigenvalues of the ground state and the first
excited state, $E^n$ ($n=1,2$), and then convert them into the
scattering phases. More ambitiously, constructing a $N \times N$
matrix allows us to look at even higher excited states. However, the
realistic computation of the scattering phase at higher energy,
$E^n$ ($n>2$), remains a challenge due to the poor signal-to-noise
ratio and the restriction of $E^n<4m_\pi$. Another way to determine
the scattering phase at more energies is to perform a lattice
calculation in the MF. In our case, we use a MF with a total
momentum $\vec{P}=\vec{e}_32\pi/L$, which provides us another two
scattering phase points. In principle, by performing the lattice
calculation in the MF with other total momenta, for example
$\vec{P}=(\vec{e}_1+\vec{e}_2)2\pi/L$, it is possible for us to
collect even more points. However, we must be careful in choosing
the MF because in some MFs the ground state and the first excited
state are nearly degenerate and isolating them becomes very
difficult.

\begin{figure}[htb]
\begin{minipage}{210pt}
\hspace{0pt}\includegraphics[width=\mywidth,angle=\plotangle]{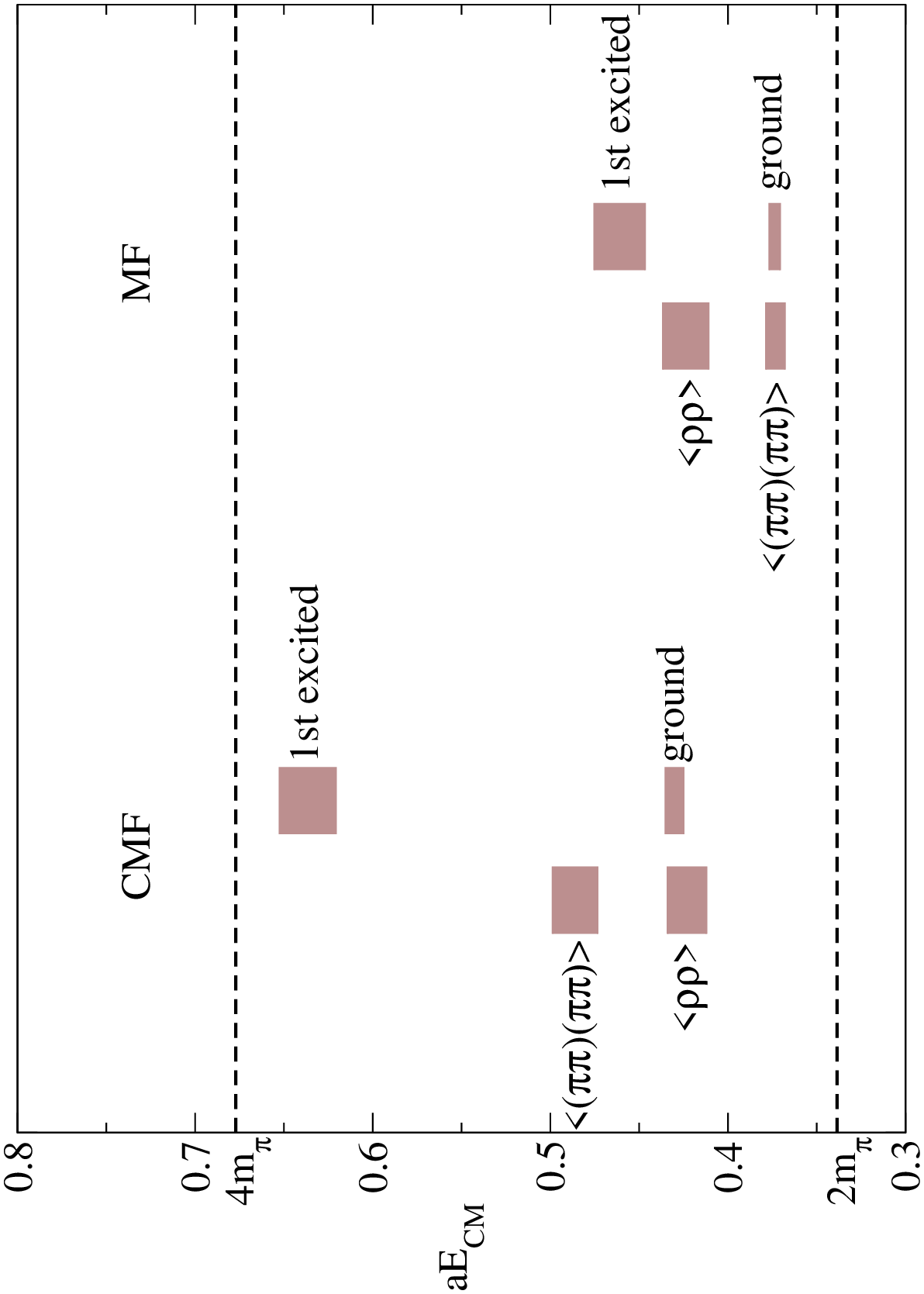}
\caption{Center-of-mass energies in the CMF and MF. The energies
evaluated from the correlation matrix are compared with the ones
evaluated from the diagonal matrix elements.}
\label{fig:invariant_mass}
\end{minipage}
\hspace{4pt}
\begin{minipage}{210pt}
\hspace{0pt}\includegraphics[width=\mywidth,angle=\plotangle]{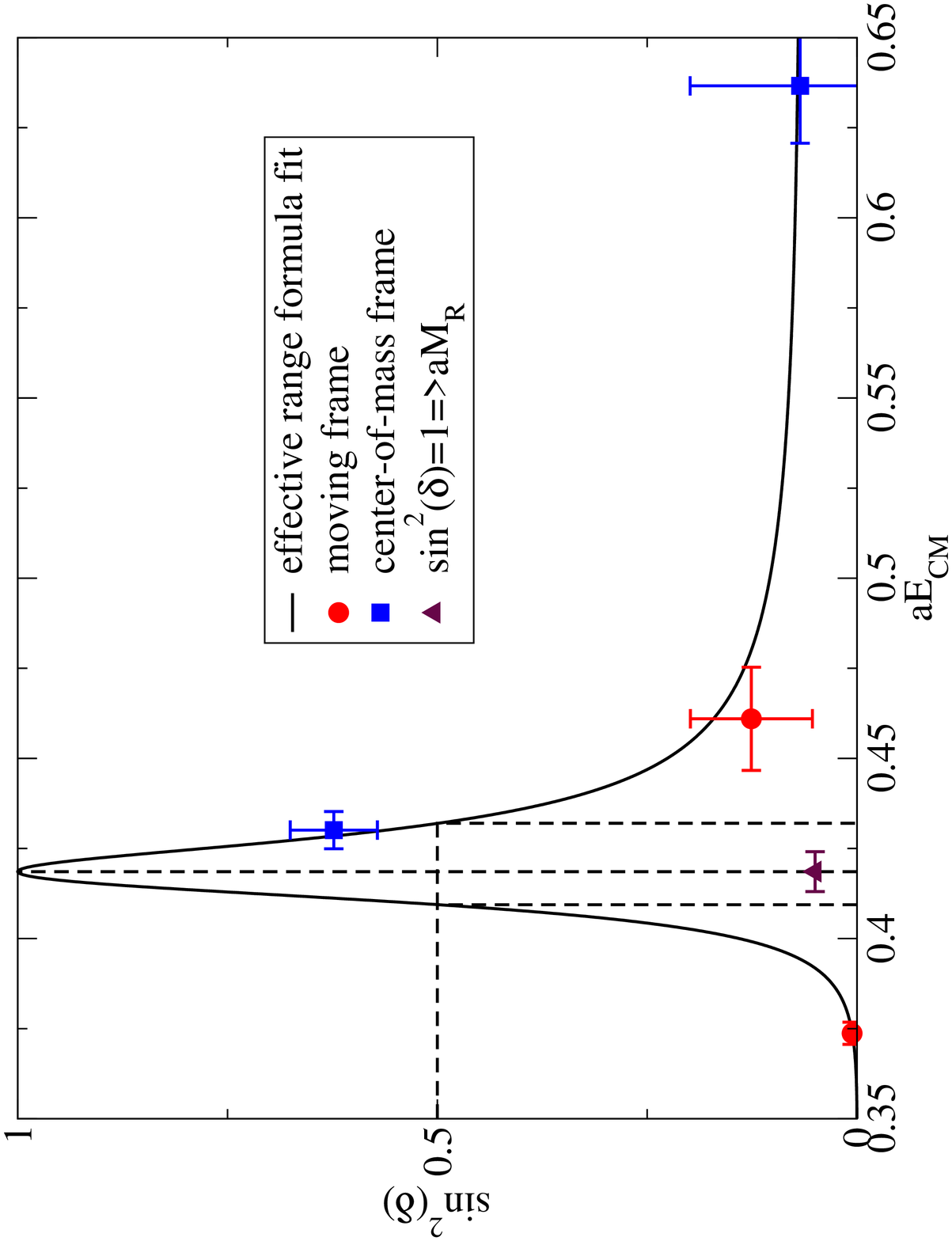}
\caption{Four scattering phases calculated on the lattice together
with the effective range formula fit. At the position where the
scattering phase passes $\pi/2$, the resonance mass $M_R$ is
determined.} \label{fig:scattering_phase_fit}
\end{minipage}
\end{figure}

In this work, both the calculations in the CMF and MF are performed
by using $N_f=2$ dynamical maximally twisted mass fermions. The
corresponding lattice parameters are $m_\pi=391~\mev$,
$a=0.086~\fm$, $L=2.1~\fm$ and $m_\pi/m_\rho=0.4$. By diagonalizing
the $2\times2$ matrix in~\ref{eq:correltaion_matrix}, the $E_{CM}$
of the ground state and the first excited state are evaluated and
shown in \Fig{\ref{fig:invariant_mass}}. In order to investigate the
effect of diagonalization, we also perform a study that utilizes
only the diagonal matrix elements. 
We see that in the CMF, there is a strong mixing
between the ground state and the first excited state in $\langle
(\pi\pi)^\dagger(t)(\pi\pi)(0)\rangle$. While in the MF, a similar
situation happens to $\langle\rho^\dagger(t)\rho(0)\rangle$. So no
operator safely provides us the ground state energy in both frames.
Therefore, introducing the diagonalization method to the calculation
of rho decay becomes essential.

\begin{figure}[htb]
\begin{minipage}{210pt}
\hspace{0pt}\includegraphics[width=\mywidth,angle=\plotangle]{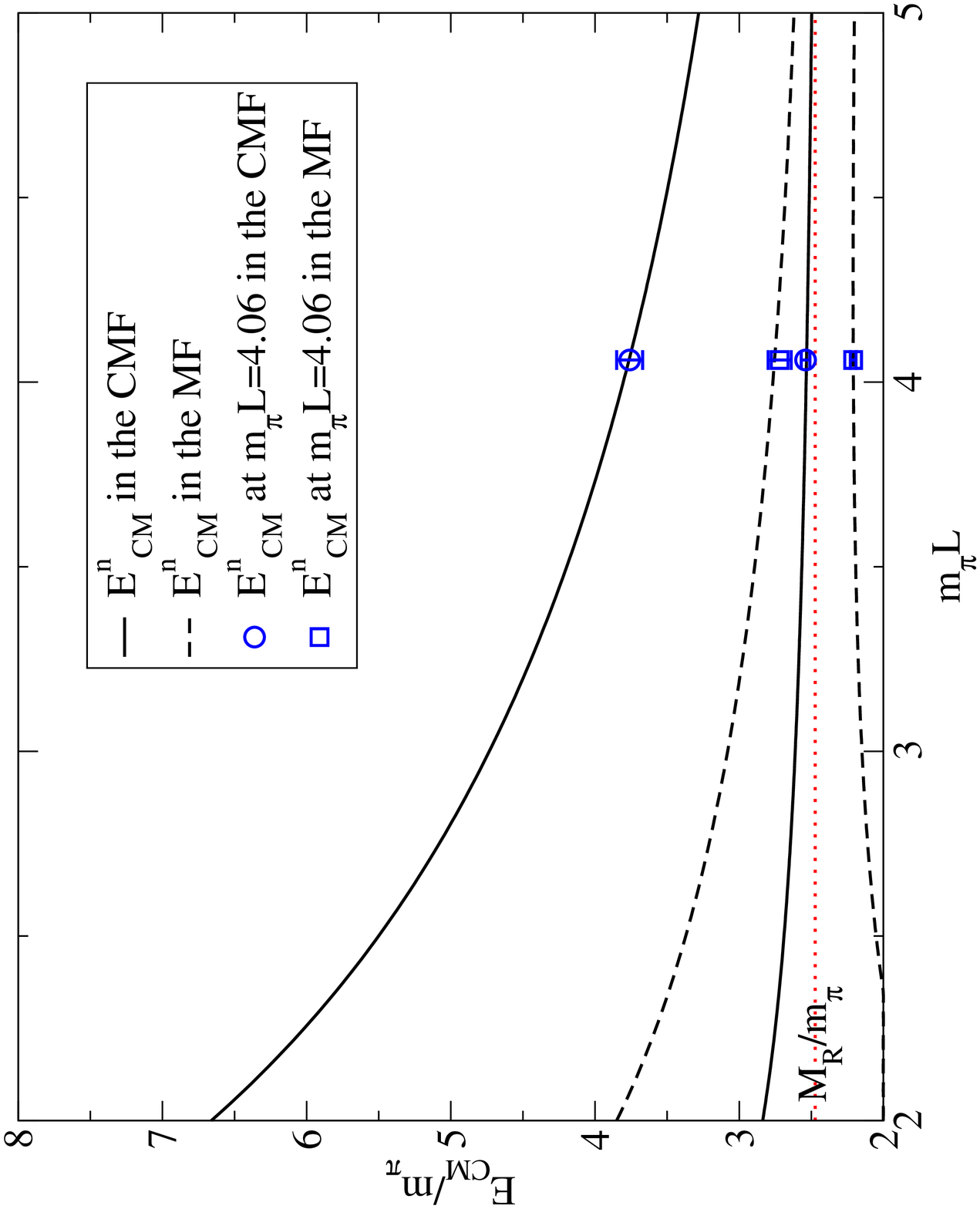}
 \caption[]{$L$ dependence of $E_{CM}^n$ (n=1,2) at a pion mass of $m_\pi=391~\mev$, 
using values of $M_R$ and $g_{\rho\pi\pi}$ in \Eq{\ref{eq:parameter}}.}
 \label{fig:expected_energy}
\end{minipage}
\hspace{4pt}
\begin{minipage}{210pt}
\hspace{0pt}\includegraphics[width=\mywidth,angle=\plotangle]{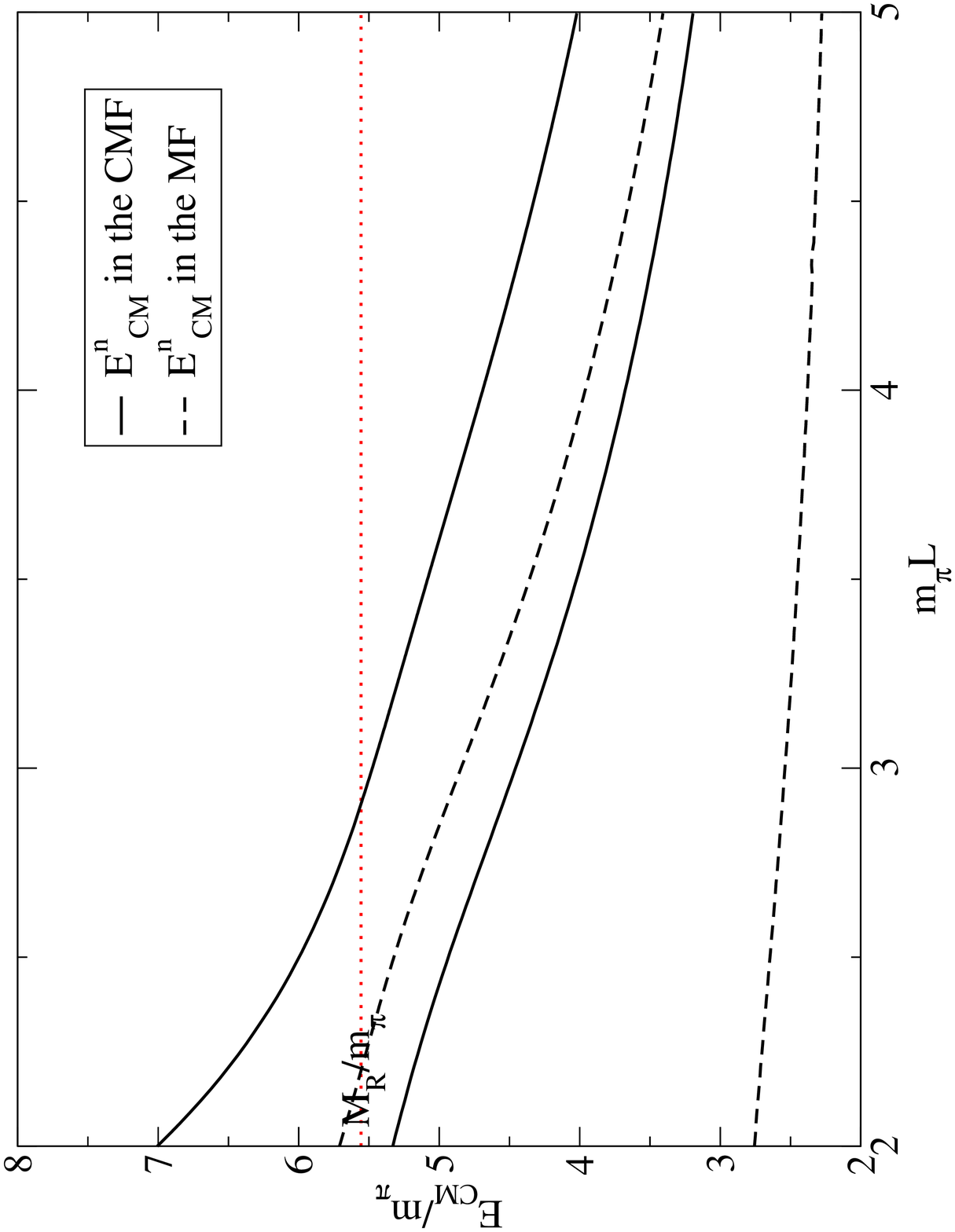}
\caption{$L$ dependence of $E_{CM}^n$ (n=1,2) in the physical limit, defined by the 
latest PDG~\cite{Amsler:2008zzb} values of $M_R$ and $g_{\rho\pi\pi}$.} 
\label{fig:expected_energy1}
\end{minipage}
\end{figure}

As shown in \Fig{\ref{fig:invariant_mass}}, all the four values of
$E_{CM}$ are smaller than $4m_\pi$. Unlike continuum QCD, twisted
mass LQCD violates the symmetries of isospin and parity. As a
result, it is possible for the rho to decay into three pions, which
means that at non-zero lattice spacing the upper bound of the
elastic scattering region is lowered to $3m_\pi$. Additionally, the
isospin symmetry breaking causes a mixing between the $I=1$ channel
and the possible $I=0$ and $I=2$ channels. Thus, a corresponding
modification would be required to adapt L\"uscher's method to the isospin
mixing case. In \Ref{\cite{Feng:2009ij}}, a significant effort was
made by us to attempt to find these effects in the $I=2$ channel, but no
compelling evidence was found.  However, the effects of isospin
violation in the I=1 channel are expected to be more
severe~\cite{Buchoff:2008hh}.  Just for the purposes of these
proceedings, we assume that such effects are small. Therefore, we
convert all the four $E_{CM}$ values into the corresponding
scattering phases using the normal method. As in
\Ref{\cite{Feng:2009ij}}, we will eventually examine the effects of
parity breaking carefully and complement the current calculation
with a calculation at a finer lattice spacing to explicitly check
for any strong lattice artifacts for $I=1$.

The results for the scattering phase are shown in
\Fig{\ref{fig:scattering_phase_fit}} together with the effective
range formula fit. At the position where the scattering phase passes
$\pi/2$, the resonance mass $M_R$ is determined. Additionally, the
values of $g_{\rho\pi\pi}$ and $\Gamma_R$ are also evaluated from
the fit. Our final results are
 \ba
 \label{eq:parameter}
 aM_R=0.4186(56)\;,\quad g_{\rho\pi\pi}=6.16(48)\quad\textmd{and}\quad a\Gamma_R=0.0217(44)\;.
 \ea
Here, our result for $g_{\rho\pi\pi}$ at $m_\pi=391~\mev$ agrees
statistically with that at the physical pion mass, which hints that
the pion mass dependence of $g_{\rho\pi\pi}$ might be weak. However,
we can not make any strong statements here since our calculation of
rho decay is only performed at one pion mass and our errors are rather large. 
To determine the chiral limit of $g_{\rho\pi\pi}$ and $M_R$, we will perform our
calculation at smaller pion masses.

Usually one takes the ground state energy $E_{CM}^{1}$ in the CMF
as the rho mass. This is correct in the case $m_\pi \geq m_\rho/2$
where the rho is still a stable particle and $E_{CM}^1$ has only an 
exponentially suppressed $L$ dependence.
However, when $m_\pi$ becomes smaller the rho becomes unstable and
$E_{CM}^{1}$ begins to depend more strongly on $L$. As an example, 
we calculate
$E_{CM}(L)$ by combining L\"uscher's formula and the effective range
formula with the parameters $M_R$ and $g_{\rho\pi\pi}$ given 
in \Eq{\ref{eq:parameter}}. As shown in
\Fig{\ref{fig:expected_energy}}, the $L$ dependence of the lowest
level is visible but still weak at a pion mass of $m_\pi=391~\mev$.
Decreasing $m_\pi$ further, more phase space becomes available for
the rho to decay into two pions. Thus, assuming $g_{\rho\pi\pi}$ is 
roughly constant, the decay width becomes larger
and the lowest level in the CMF begins to look more like a scattering state.
\Fig{\ref{fig:expected_energy1}} shows $E_{CM}(L)$ in
the extreme limit at the physical pion mass. There we see that
$E_{CM}^1$ drops so rapidly with $L$ that it bears no resemblance to a stable 
state with mass $M_R$.

\section{Conclusions}
\label{sect:conclusion} We have calculated the S-wave pion-pion
scattering length in the $I=2$ channel and the P-wave pion-pion
scattering phase in the $I=1$ channel using $N_f=2$ maximally
twisted mass fermions. In the former channel, the pion masses ranged
from $270~\mev$ to $485~\mev$. Using \chipt{} at NLO, we
extrapolated our results for the scattering length to the physical
limit, where we found $m_\pi \atwo=-0.04385\,(28)(38)$ and
$l_{\pi\pi}^{I=2}(\mu=\fphy)=4.65\,(.85)(1.07)$.
In the latter channel, we performed a calculation at
$m_\pi=391~\mev$, $a=0.086~\fm$ and $m_\pi/m_\rho=0.4$.
Making use of finite size methods, we evaluated the scattering phase
at four energies, to which we fit the effective range formula and
found the results $aM_R=0.4186(56)$, $g_{\rho\pi\pi}=6.16(48)$ and
$a\Gamma_R=0.0217(44)$.

\section{Acknowledgments}
This work was supported by the DFG project Mu 757/13 and the DFG
Sonderforschungsbereich / Transregio SFB/TR9-03. We thank
G.~Herdoiza, A.~Shindler, C.~Urbach and M.~Wagner for valuable
suggestions and assistance. X.~Feng would like to thank C.~Liu,
G.~M\"unster, N.~Ishizuka and A.~Walker-Loud for helpful
communications. The computer time for this project was made
available by the John von Neumann Institute for Computing on the
JUMP and JUGENE systems in J\"ulich. We also thank the staff of the
computer center in Zeuthen for their technical support.

\bibliography{pipi}

\end{document}